\PassOptionsToPackage{usenames}{xcolor}
\PassOptionsToPackage{dvipsnames}{xcolor}

\documentclass[sigconf]{acmart}

\usepackage{subcaption}
\usepackage{enumitem}
\usepackage{amsmath}
\usepackage{url}
\usepackage{balance}
\usepackage{siunitx}
\usepackage[usenames]{xcolor}

\usepackage{array}
\newcolumntype{P}[1]{>{\centering\arraybackslash}m{#1}}
\newcolumntype{L}[1]{>{\arraybackslash}m{#1}}
\usepackage{xspace}
\newcommand\ie{i.\,e.\xspace}
\newcommand\eg{e.\,g.\xspace}

\newcommand\US{U.\,S.\xspace}

\newcommand{\mathup}[1]{\text{\textup{#1}}}

\AtBeginDocument{%
  \providecommand\BibTeX{{%
    \normalfont B\kern-0.5em{\scshape i\kern-0.25em b}\kern-0.8em\TeX}}}

\copyrightyear{2023}
\acmYear{2023}
\setcopyright{acmlicensed}
\acmConference[WWW '23]{Proceedings of the ACM
Web Conference 2023}{May 1--5, 2023}{Austin, TX, USA}
\acmBooktitle{Proceedings of the ACM Web Conference 2023 (WWW '23), May
1--5, 2023, Austin, TX, USA}
\acmDOI{10.1145/3543507.3583857}
\acmISBN{}

\begin{document}

\fancyhead{}

\title{Believability and Harmfulness Shape the Virality of\\Misleading Social Media Posts}

\author{Chiara Drolsbach}
\email{chiara.drolsbach@wi.jlug.de}
\affiliation{
	\institution{JLU Giessen}
	\streetaddress{Licher Str.\ 74}
	\country{Germany}
}

\author{Nicolas Pröllochs}
\email{nicolas.proellochs@wi.jlug.de}
\affiliation{
	\institution{JLU Giessen}
	\streetaddress{Licher Str.\ 74}
	\country{Germany}
}
\renewcommand{\shortauthors}{Chiara Drolsbach and Nicolas Pröllochs}

\begin{abstract}
Misinformation on social media presents a major threat to modern societies. While previous research has analyzed the virality across true and false social media posts, not every misleading post is necessarily equally viral. Rather, misinformation has different characteristics and varies in terms of its believability and harmfulness -- which might influence its spread. 
In this work, we study how the perceived believability and harmfulness of misleading posts are associated with their virality on social media. Specifically, we analyze (and validate) a large sample of crowd-annotated social media posts from Twitter's Birdwatch platform, on which users can rate the believability and harmfulness of misleading tweets. To address our research questions, we implement an explanatory regression model and link the crowd ratings for believability and harmfulness to the virality of misleading posts on Twitter. Our findings imply that misinformation that is (i) easily believable and (ii) not particularly harmful is associated with more viral resharing cascades. These results offer insights into how different kinds of crowd fact-checked misinformation spreads and suggest that the most viral misleading posts are often not the ones that are particularly concerning from the perspective of public safety. From a practical view, our findings may help platforms to develop more effective strategies to curb the proliferation of misleading posts on social media.
\end{abstract}




\maketitle

\section{Introduction}

Social media disseminates vast amounts of misinformation \cite[\eg,][]{Vosoughi.2018, Shao.2016,Ecker.2022}. Several works have studied the diffusion of rumors of varying veracity, finding that misinformation spreads more virally than the truth \cite{Friggeri.2014,Vosoughi.2018,Prollochs.2021b,Solovev.2022b,Bessi.2015}. If misinformation becomes viral, it can have detrimental real-world consequences and affects how opinions are formed \cite{Allcott.2017,Bakshy.2015,DelVicario.2016,Oh.2013}. This has been observed, for example, during elections \cite[\eg,][]{Aral.2019,Allcott.2017,Bakshy.2015,Grinberg.2019} and crisis situations \cite[\eg,][]{Pennycook.2020b,Solovev.2022b,Oh.2013,Oh.2010,Starbird.2014,Broniatowski.2018,Zeng.2016,Geissler.2022}. 
As such, misinformation on social media threatens the well-being of society at large and demands effective countermeasures \cite{Lazer.2018,Bar.2023,Pennycook.2021}. 

While earlier research has analyzed differences in the spread of true and false social media posts \cite{Vosoughi.2018,Prollochs.2021b,Solovev.2022b}, not every misinforming post is necessarily equally viral. Rather, misinformation has different characteristics and varies in terms of its believability and harmfulness -- which might influence its spread. For example, individuals using social media tend to be in a hedonic mindset and thus are looking for entertainment and fun \cite{Kim.2019,Lutz.2020}. Thus, if a user does not believe the content of a post, there might be less incentive to share it and increase its reach. 
In a similar vein, research in psychology suggests that threats capture attention \cite{Schmidt.2015, Koster.2004}. Contextualized to misinformation on social media, this would imply that harmful misleading posts are detected more accurately -- and, therefore, less likely to be shared. Overall, one may expect that the believability and harmfulness of misinformation play a crucial role in its spread. However, there is currently no study empirically analyzing the link between these attributes and virality on social media. 

\textbf{Research goal:}
We analyze the link between the believability and harmfulness of misleading posts and their virality on social media. In particular, we seek to answer two research questions:
\begin{itemize}
\item \textbf{(RQ1)} \emph{Are misleading posts perceived as believable more viral than those perceived as not believable?}
\item \textbf{(RQ2)} \emph{Are misleading posts perceived as harmful more viral than those perceived as not harmful?}
\end{itemize}

\textbf{Data \& methods:} 
We draw upon a large dataset of crowd-annotated tweets from Twitter's fact-checking system ``Birdwatch'' \cite{Prollochs.2022a}. On Birdwatch, users can create ``Birdwatch notes'' that aim to identify misleading tweets directly on Twitter. A unique feature of fact-checking on Birdwatch is that users also categorize whether they perceive misleading tweets to be easily believable and/or harmful. For our analysis, we collect (and validate) Birdwatch notes for misleading tweets between the launch of Birdwatch in early 2021 and the end of February 2022. Subsequently, we perform an explanatory regression analysis and link the believability and harmfulness (as provided in Birdwatch notes) to the number of retweets (as a measure of virality) of the fact-checked post. In our analysis, we control for established predictors that may affect the retweet rate (\eg, social influence, sentiment). This approach allows us to empirically test how the believability and harmfulness of misleading posts are associated with their virality on social media. 

\textbf{Contributions:} Our study offers insights into how crowd fact-checked misinformation spreads on social media. Specifically, we demonstrate that misinformation that is (i) easily believable and (ii) not particularly harmful is associated with more viral resharing cascades. These findings imply that not all kinds of misinformation are equally viral; and that the most viral misleading posts are oftentimes not the ones that are particularly concerning from the perspective of public safety. In a next step, our findings may help platforms to implement more effective strategies for reducing the proliferation of misinformation.

\section{Background}
\label{sec:related_work}

\textbf{Community-based fact-checking:} 
The concept of community-based fact-checking is a relatively novel approach that aims to tackle misinformation on social media by harnessing the ``wisdom of crowds'' \cite{Woolley.2010,Frey.2021}. Specifically, the idea is to let regular social media users carry out fact-checking of social media posts \cite{Micallef.2020,Bhuiyan.2020,Pennycook.2019,Epstein.2020,Allen.2020,Allen.2021,Godel.2021}. Compared to expert-based approaches to fact-checking (e.g., via third-party fact-checking organizations), community-based fact-checking is appealing as it allows for large numbers of fact-checks to be frequently and inexpensively acquired \cite{Woolley.2010,Allen.2021}. Moreover, it addresses the issue that many users do not trust the assessments of professional fact-checkers (\eg, due to alleged political biases) \cite{Poynter.2019}. Experimental studies suggest that the crowd can be highly accurate in identifying misinformation and even relatively small crowds can yield performance similar to experts \cite{Epstein.2020,Pennycook.2019,Bhuiyan.2020}. 

\textbf{Birdwatch:} Informed by experimental studies, the social media platform Twitter has recently launched its community-based fact-checking system Birdwatch \cite{Twitter.2021,Prollochs.2022a}. Different from earlier crowd-based fact-checking initiatives \cite{Oriordan.2019,Florin.2010,Bhuiyan.2020,Bakabar.2018}, Birdwatch allows users to identify misinformation \emph{directly} on the platform (see next section for details). Given the recency  of the platform, research on Birdwatch is scant. Early works suggest that politically motivated reasoning might pose challenges in community-based fact-checking \cite{Allen.2022,Prollochs.2022a}. Notwithstanding, community-created fact-checks on Birdwatch have been found to be perceived as informative and helpful by the vast majority of social media users \cite{Prollochs.2022a}. Furthermore, real-world community fact-checks have been shown to be effective in reducing users' propensity to reshare misinformation \cite{Wojcik.2022}.

\textbf{Virality of misinformation:} 
Several works have analyzed the spread of social media posts for which veracity was determined based on the assessment of third-party fact-checking organizations \cite{Friggeri.2014,Vosoughi.2018,Solovev.2022b,Prollochs.2021b,Prollochs.2022b}. For instance, \citet{Friggeri.2014} analyzed upload and deletion rates in $\approx$4,000 expert fact-checked rumors from Facebook. Another literature stream has analyzed the diffusion of true vs. false rumors on Twitter \cite{Vosoughi.2018,Solovev.2022b,Prollochs.2021b,Prollochs.2022b}. The rumors (and their veracity) in these works were identified based on the presence of user comments referencing fact-checks carried out by third-party fact-checking organizations (see, \eg, \citet{Vosoughi.2018} for methodological details). These studies typically observed that false social media posts spread more viral than true posts. 

\textbf{Research gap:} Existing research has primarily focused on studying the virality across true vs. false social media posts that have been fact-checked by expert fact-checkers. However, an understanding of how the virality of misinformation varies depending on its underlying characteristics is largely absent. Specifically, we are not aware of previous work empirically analyzing how the perceived believability and harmfulness of misleading posts are associated with their virality on social media. This presents our contribution.

\section{Data and Methodology}
\label{sec:methods}

\subsection{Data Collection}

To answer our research questions, we analyze a large dataset of crowd-annotated tweets that have been identified as being misleading during the pilot phase of Twitter's Birdwatch platform \cite{Twitter.2021,Prollochs.2022a}. Birdwatch has been launched by Twitter on January 23, 2021, and aims to identify misleading social media posts by harnessing the wisdom of crowds. Different from earlier small-scale crowd-based initiatives to fact-checking~\cite{Oriordan.2019,Florin.2010,Bhuiyan.2020,Bakabar.2018}, Birdwatch allows users to identify misleading tweets \emph{directly} on Twitter and write short (max 280 characters) fact-checks (so-called ``Birdwatch notes'') that add context to the tweet. Another unique feature of Birdwatch is that authors of Birdwatch notes additionally need to answer checkbox questions when identifying misleading posts. Here users can rate whether they perceive the misleading tweet to be easily believable and whether the tweet might cause considerable harm. 

\begin{figure}[h]
	\centering
	{\fbox{\includegraphics[width=0.475\columnwidth]{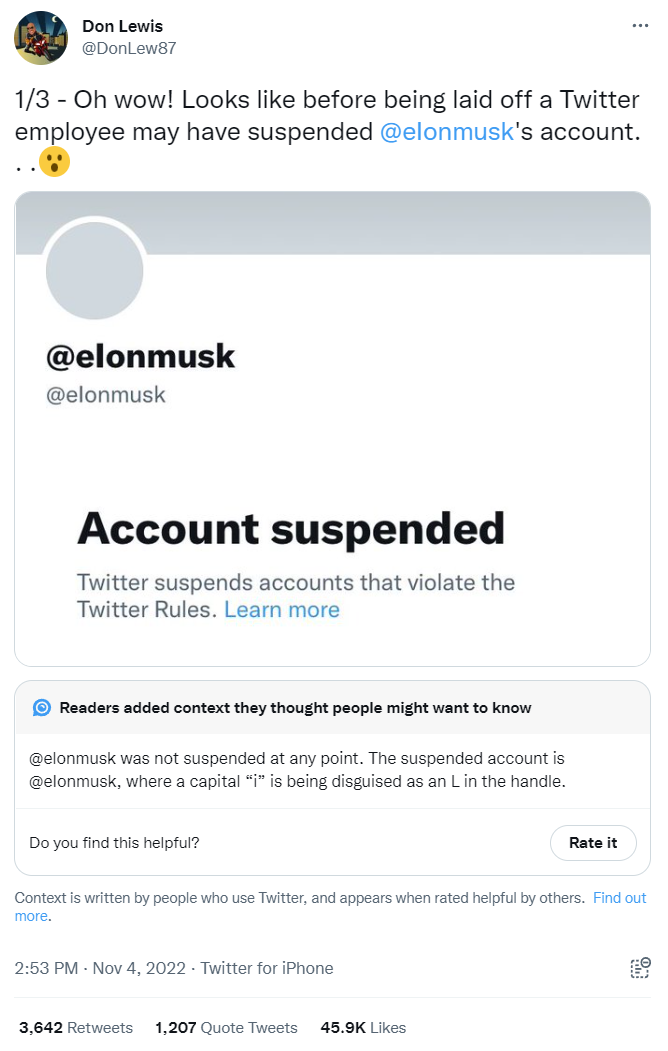}}}
	\caption{Example of a community note (\ie, Birdwatch note) identifying a misleading post on Twitter.}
	\label{fig:screenshot} 
\end{figure}

To participate in the pilot phase of the Birdwatch feature (only available in the US), Twitter users had to register and apply to become a contributor. In early 2022, Birdwatch had approximately 3250 contributors, which is a relatively small fraction of all Twitter users ($\approx$41.5 million daily active users \cite{Statista.2022}). Birdwatch notes were displayed directly on tweets to pilot participants (see example in Fig.~\ref{fig:screenshot}); while all other Twitter users could view them on a separate Birdwatch website (\url{birdwatch.twitter.com}). Accordingly, the fact-checks were not directly visible to the vast majority of Twitter users. Birdwatch notes were thus unlikely to influence the diffusion of the fact-checked tweets during our study period.

For our analysis, we downloaded {all} Birdwatch notes between the launch of Birdwatch on January 23, 2021, and the end of February 2022 from the Birdwatch website\footnote{Available via \url{https://twitter.com/i/communitynotes/download-data}.}, \ie, for an observation period of more than one year. The dataset contains a total number of \num{20218} Birdwatch notes from \num{3257} different contributors. 

On Birdwatch, multiple users can write Birdwatch notes for the same tweet. Therefore, the data sometimes includes multiple Birdwatch notes for the same post ($\approx$ 1.24 notes per tweet). As a result, different Birdwatch users might disagree on the characteristics of one tweet. To incorporate this, we used majority vote to determine the categorizations. We excluded tweets without a definite assessment (\ie, if two assessments stand in opposition) and tweets classified as not misleading.\footnote{Birdwatch contributors can also endorse the accuracy of \emph{not} misleading tweets (\SI{5.72}{\percent} of all Birdwatch notes). Since users cannot rate the believability and harmfulness of these tweets, Birdwatch notes for not misleading tweets are excluded from our analysis.} This filtering step resulted in a dataset consisting of \num{13732} tweets. Each of the fact-checks addresses a single \emph{misleading} tweet for which the Birdwatch contributor has assessed the believability and harmfulness.

We further mapped the \emph{tweetID} referenced in each Birdwatch note to the underlying source tweets using the Twitter historical API. This allowed us to collect additional information concerning the fact-checked tweets and its author, namely, (a) the number of retweets, (b) the followers count, (c) the followees count, (d) the account age, and (e) whether the user has been verified by Twitter. Moreover, we calculated a sentiment score for each source tweet to control for its positivity/negativity in our later empirical analysis.\footnote{Analogous to prior work \cite[\eg,][]{Prollochs.2021b,Jakubik.2023,Upworthy.2023}, we use the NRC dictionary \cite{Mohammad.2013} to calculate a sentiment score measuring the share of positive vs. negative words. Here, we use the default implementation for sentiment analysis provided in the \texttt{sentimentr} R package.}

\subsection{Explanatory Regression Model}

We specify an explanatory regression model that explains the virality of misleading tweets based on their believability and harmfulness. In our analysis, we use a common proxy for the virality of a resharing cascade, namely, the number of retweets \cite{Han.2020,Solovev.2022b}. Since the variance of the retweet count is larger than its mean, we have to adjust for overdispersion. Analogous to earlier research \cite[\eg,][]{Stieglitz.2013,Solovev.2022b}, we thus employ a negative binomial regression model. 

Formally, the response variable in our negative binomial regression model is $\textit{RetweetCount}_i$, which refers to the number of retweets received by tweet $i$. The key regressors are binary and indicate whether the tweet has been rated as believable ($\mathit{Believable}_i$, $=1$ if true, otherwise 0) and harmful ($\mathit{Harmful}_i$, $=1$ if true, otherwise 0) on Birdwatch. Concordant with earlier work \cite[eg,][]{Stieglitz.2013,Solovev.2022b,Vosoughi.2018}, we control for the social influence of the author of the source tweet (\eg, some authors have many followers and reach wider audiences). The control variables comprise the followers and followees count, the account age (in years), and the verification status. Furthermore, we control for the positivity/negativity ($Sentiment$) of the fact-checked tweet. This yields the model
{
	\begin{align}
		\log&({\mathup{E}(RetweetCount_i \,\mid\, ^*)})  = \,\beta_0  + \beta_{1} \, \mathit{Believable}_i +  \beta_{2} \, \mathit{Harmful}_i \nonumber  \\
		& + \beta_{4} \,  \mathit{Sentiment}_i  + \beta_{5} \, \mathit{Followers}_i + \beta_{6} \, \mathit{Followees}_i  \nonumber \\ 
		& + \beta_{7} \, \mathit{Account Age}_i + \beta_{8} \, \mathit{Verified}_i + u_{i}, \label{eq:neg_bin_misleading} 
	\end{align} 
}%
with intercept $\beta_0$. Furthermore, we include month-year fixed effects $u_{i}$, which allow us to control for varying start dates and the age of the resharing cascades \cite[\eg,][]{Solovev.2022b}. In our regression analysis, all continuous variables are $z$-standardized to facilitate interpretability. 

\section{Empirical Analysis}

\subsection{Summary Statistics}

\begin{figure}[h]
	\captionsetup{position=top}
	\centering
	\subfloat[]{\includegraphics[width=.475\linewidth]{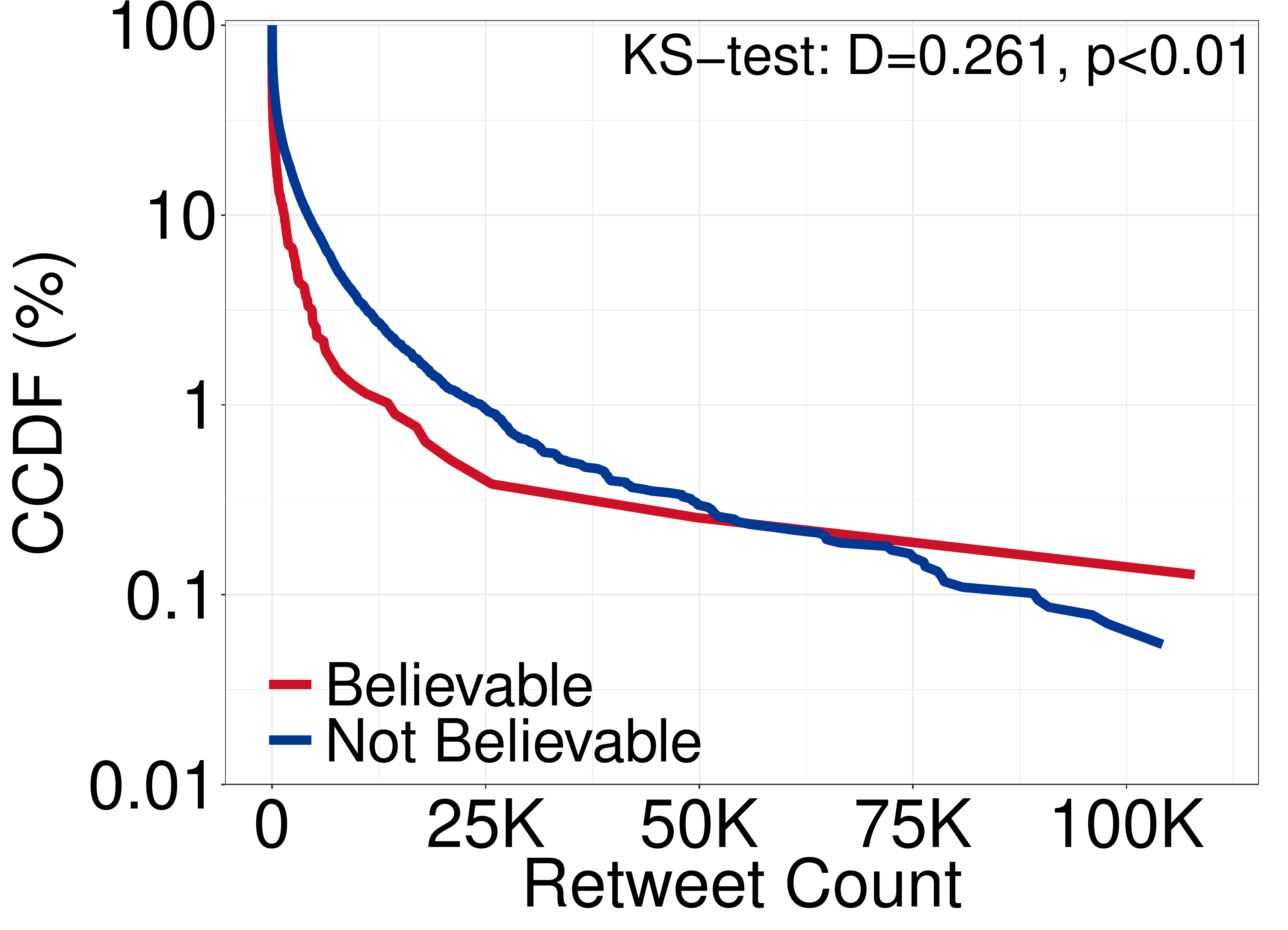}\label{fig:ccdf_believability}}
	\hspace{0.2cm}
	\subfloat[]{\includegraphics[width=.475\linewidth]{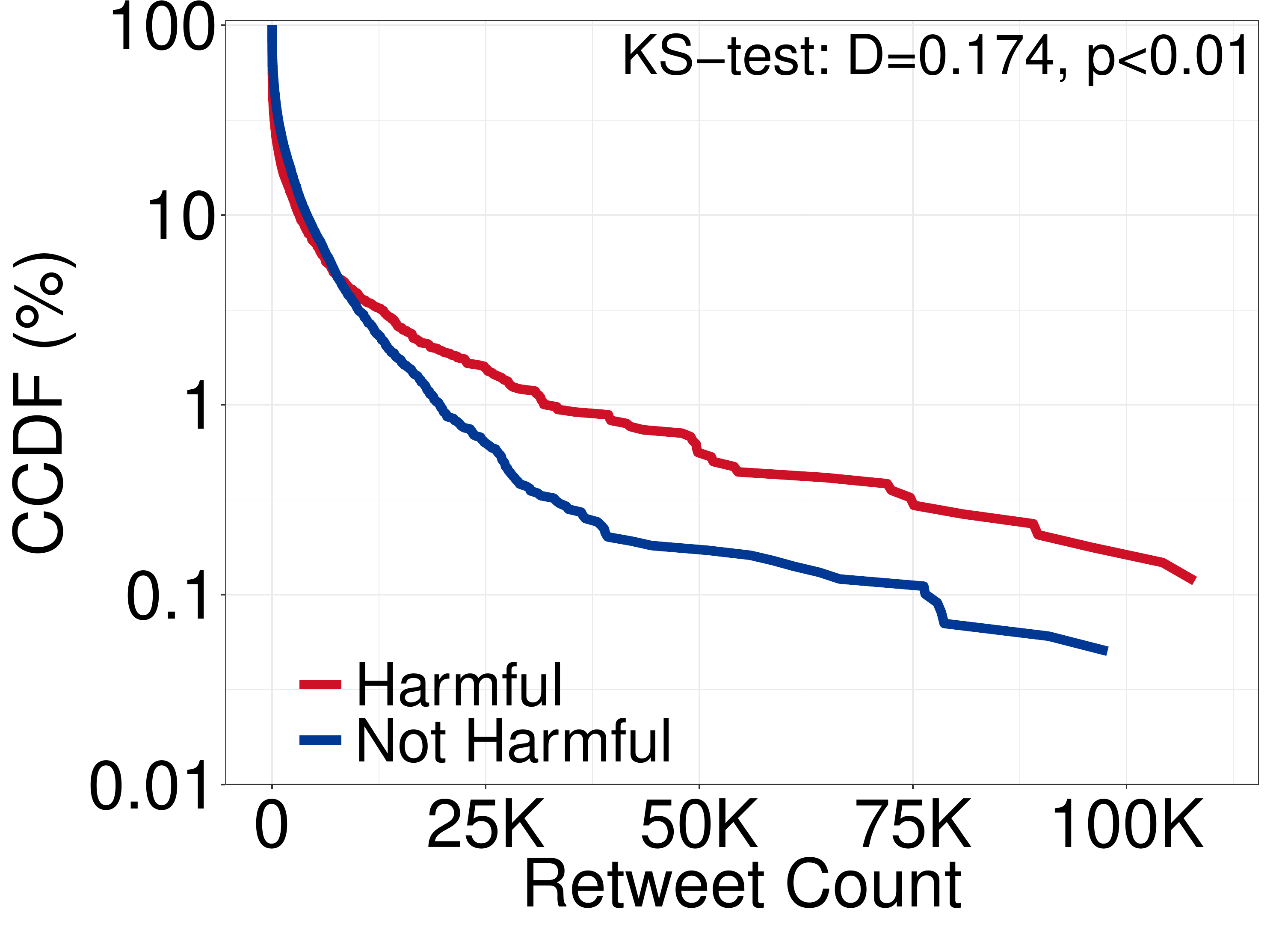}\label{fig:ccdf_difficulty}}
	\vspace{.25cm}
	\caption{Complementary cumulative distribution functions showing the distribution of the retweet count separated by (a) believability and (b) harmfulness.}
	\label{fig:ccdfs}
\end{figure}

We start our analysis by evaluating summary statistics. Out of all tweets, \SI{94.20}{\percent} are rated as $\mathit{Believable}$ and \SI{74.60}{\percent} as $\mathit{Harmful}$. In total, the tweets have received \num{26.81} million retweets. On average, each tweet in our dataset has received \num{1724} retweets. However, the number of retweets is higher for tweets perceived as believable. Specifically, the average number of retweets is \num{1772} for believable tweets and \num{751} for not believable tweets. We further observe that tweets rated as harmful receive fewer retweets (\num{1607}) than tweets rated as not harmful (\num{1832}). Complementary cumulative distribution functions for the retweet count are shown in Fig.~\ref{fig:ccdfs}. The differences in the distributions are statistically significant according to two-tailed Kolmogorov-Smirnov (KS) tests  ($p<0.01$). Additionally, we calculated the correlation between the variables $\mathit{Believable}$ and $\mathit{Harmful}$. Here we find a weak positive correlation of {\num{0.181} ($p<0.01$)}. This indicates that harmful posts can be but are not necessarily believable (and vice versa).

Note that the tweets in our dataset show substantial heterogeneity regarding the characteristics of the source accounts. On average, the authors of the tweets have \num{1.39} million followers (SD: \num{5.88} million), \num{5795} followees (SD: \num{20094}), and an account age of \num{8.89} years (SD: \num{4.46}). A total share of \SI{47.90}{\percent} of all authors have been verified by Twitter (SD: \num{0.50}). The mean sentiment of the tweets in our dataset is \num{-0.005}, \ie, slightly negative (SD: \num{0.26}). To accommodate these potentially confounding factors, we estimate an explanatory regression model with control variables in the next section.

\subsection{Regression Analysis}

\begin{figure}[h]
	\centering
	\includegraphics[width=.825\columnwidth]{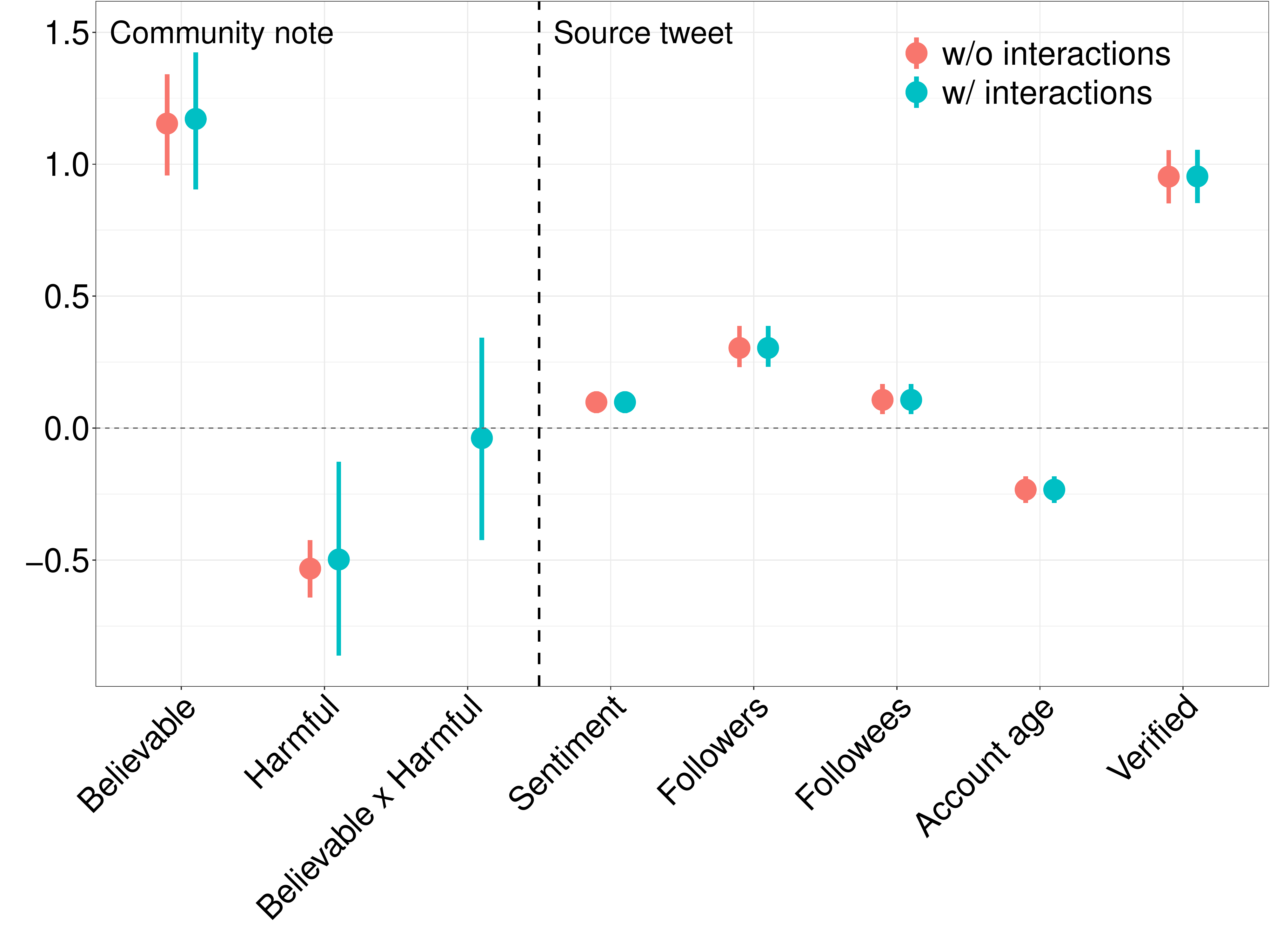}
	\vspace*{-5mm}
	\caption{Negative binomial regression linking perceived believability and harmfulness to the number of retweets. Reported are models w/o (coral) and w/ (turquoise) an interaction term between believability and harmfulness. The circles show standardized coefficient estimates and the error bars indicate the \SI{99}{\percent} CIs. Month-year fixed effects are included. 
	} 
	\label{fig:regression}
\end{figure}

\textbf{Coefficient estimates:} We estimate a negative binomial regression to study the role of believability and harmfulness in the virality of misleading posts after controlling for confounding effects (\eg, varying social influence). 
Fig.~\ref{fig:regression} reports the coefficient estimates and 99\% CIs. The dependent variable is the retweet count of the misleading tweet. We find that the coefficient for $\mathit{Believable}$ is positive and statistically significant (coef: $1.154$; $p<0.01$). This estimate implies that misleading posts perceived as believable receive $e^{1.154}-1 \approx$ \SI{217.09}{\percent} more retweets. We further observe a negative and statistically significant coefficient for $\mathit{Harmful}$ (coef: $-0.533$; $p<0.01$). This implies that misleading posts perceived as harmful receive \SI{41.32}{\percent} fewer retweets. In sum, we find that misinformation that is (i) easily believable and (ii) not particularly harmful is associated with more viral resharing cascades. 

\textbf{Interaction effect:} Misleading posts can be categorized as (i) believable or harmful, (ii) believable and harmful, or (iii) neither believable nor harmful. To test whether the effects of different combinations of believability and harmfulness on virality differ, we reestimated our regression model with an interaction term between $\mathit{Believable} \times \mathit{Harmful}$ (see Fig~\ref{fig:regression}). We observe that the coefficient of the interaction term is not statistically significant (coef: $0.038$; $p=0.796$). At the same time, the coefficients of $\mathit{Believable}$ and $\mathit{Harmful}$ remain stable. This suggests that the predictors' effects are additive and do not depend on each other. 

\textbf{Control variables:} We also observe statistically significant coefficient estimates for the control variables in our regression analysis. Specifically, more retweets occur for source tweets authored by accounts with higher numbers of followers (coef: $0.303$; $p<0.01$) and followees (coef: $0.107$; $p<0.01$). Furthermore, more retweets are estimated for tweets from accounts that are younger in age (coef: $-0.233$; $p<0.01$) and users with a verified status (coef: $0.953$; $p<0.01$). Analogous to prior work \cite[\eg,][]{Prollochs.2021a,Prollochs.2021b}, we also observe that resharing cascades are larger if they convey a more positive sentiment (coef: $0.098$; $p<0.01$).

\textbf{Robustness checks:} 
We carried out multiple checks that confirmed the robustness of our results. First, we checked our models for multicollinearity and ensured that the VIFs are below four. Second, we reestimated our models with a random-effects specification controlling for heteregoneity across the contributors on Birdwatch (\ie, user-specific effects). Third, we used alternative methods for handling multiple Birdwatch notes for the same source tweets (\eg, via Birdwatch's rating mechanism; see \cite{Twitter.2021,Prollochs.2022a}). In each of these checks, we found support for our findings.

\subsection{Validation Study}

To validate the categorizations on Birdwatch, we carried out a user study with \textit{n = 7} participants via Prolific (\url{www.prolific.co}). All participants were English native speakers and based in the \US Furthermore, six out of seven participants stated that they regularly use social media to share content. We asked the participants to rate the believability and harmfulness of \num{150} misleading tweets from Birdwatch on a 5-point Likert scale. The participants rated tweets categorized as believable by Birdwatch users as significantly more believable than tweets not categorized as believable ($M_{\text{Believable/Believable}}=3.61$, $M_{\text{Believable/NotBelievable}}=3.25$, $t = 3.03$, $p<0.01$). Furthermore, tweets categorized as harmful by Birdwatch users were rated as significantly more harmful than misleading tweets not categorized as harmful ($M_{\text{Harmful/Harmful}}=3.50$, $M_{\text{Harmful/NotHarmful}}=3.02$, $t = 5.30$, $p<0.01$). The inter-rater agreement was statistically significant for both believability ($W = 0.27$, $p<0.01$) and harmfulness ($W = 0.43$, $p<0.01$). These findings add to the validity of our results and confirm that the perceptions of independent annotators (that may have varying familiarity with the tweets' information) and the categorizations of (self-selected) Birdwatch users point in the same direction.

\section{Discussion}

\textbf{Research implications: }
We contribute to research into misinformation by studying the link between specific attributes of misleading posts and their virality on social media. Specifically, we hypothesized that the virality of misleading posts differs depending on the perceived (i) believability and (ii) harmfulness. Our results suggest that misleading posts that are easily believable are more viral. From a theoretical perspective, a possible explanation lies in the hedonic mindset of social media users: if a user does not believe the content of a post, increasing its reach might be less enjoyable \cite[\eg,][]{Johnson.2015,Kim.2019,Minas.2014,Moravec.2019}. We further found that misleading posts perceived as harmful are less viral than those perceived as not harmful. This finding is concordant with research in psychology \cite[\eg,][]{Koster.2004,Schmidt.2015,vanDamme.2008}, suggesting that humans are more attentive if confronted with potentially harmful information. As a result, harmful misinformation might be detected more accurately 
and, therefore, less likely to be shared. Altogether, our work provides novel insights into how community fact-checked posts spread in a real-world environment and demonstrates that not all kinds of misinformation are equally viral. While previous research \cite[\eg,][]{Vosoughi.2018,Solovev.2022b,Prollochs.2021b} has analyzed differences in the spread of rumors of varying veracity, this study is the first to empirically study how the perceived believability and harmfulness of misleading posts are linked to their virality on social media.

\textbf{Practical implications: }
Our findings are relevant for the design of more sophisticated strategies to counter misinformation. Community-based fact-checking has the potential to partially overcome the drawbacks of the experts' approach to fact-checking, \eg, in terms of speed, volume, and trust \cite{Pennycook.2019}. Our observation that viral misleading posts tend to be easily believable and not particularly harmful implies that the most viral community fact-checked misinformation is often not particularly concerning from the perspective of public safety. In practice, this knowledge could be used by platforms to enhance the prioritization of posts for expert fact-checking. Our findings may also be relevant with regard to educational applications and for enhancing the accuracy of machine learning models for automatically detecting misleading posts. 

\textbf{Limitations and future work: }
As with others, our study is not free of limitations and offers potential for future work. First, analogous to earlier observational studies \cite[\eg,][]{Vosoughi.2018,Solovev.2022b,Solovev.2023,Prollochs.2021b}, we demonstrate associations and not causal paths. Second, experimental studies in controlled settings may help to understand whether the perceptions regarding the believability and harmfulness of misinformation differ between community fact-checkers, experts, and regular social media users. Third, the restricted set of community fact-checked posts on Birdwatch may not reflect the overall population of misleading posts on social media. Thus, more research is necessary to better understand how the crowd selects posts for fact-checking \cite{Drolsbach.2022}. For instance, it would be interesting to understand whether Birdwatch users are more likely to fact-check tweets that are easier to judge in terms of their believability and harmfulness. Fourth, our analysis is limited to the social media platform Twitter and data from the Birdwatch pilot phase. In the future, community-based fact-checking on Twitter may evolve to a different steady-state due to a growing/more experienced user base and changes in functionality (\eg, Twitter recently rebranded Birdwatch to ``Community Notes'' \cite{Twitter.2021}). Fifth, future work may analyze whether the observed spreading patterns are generalizable to posts from other fact-checking systems and social media platforms. 

\section*{Acknowledgments}
This study was supported by a grant from the German Research Foundation (DFG grant 455368471). 

\bibliographystyle{ACM-Reference-Format-no-doi-abbrv}
\balance
\bibliography{literature}

\end{document}